\documentclass[useAMS,usenatbib]{mn2e}

\usepackage{graphicx}

\title[Swift observations of GW Lib]{Swift observations of GW Lib: a unique insight into a rare outburst} 
\author[K. Byckling et al.]{K. Byckling,$^{1}$\thanks{E-mail: kjkb2@star.le.ac.uk}
 J.P. Osborne,$^{1}$ P.J. Wheatley,$^{2}$ G.A. Wynn,$^{1}$ 
 A. Beardmore,$^{1}$ \and V. Braito,$^{1}$ 
 K. Mukai,$^{3}$ and R.G. West$^{1}$\\
$^{1}$Department of Physics and Astronomy, University of Leicester, University Road, Leicester, LE1 7RH, UK\\ 
$^{2}$Department of Physics, University of Warwick, Coventry, CV4 7AL, UK\\
$^{3}$NASA/Goddard Space Flight Center, Greenbelt, MD 20771, USA}
\begin{document}

\date{Accepted . Received ; in original form }

\pagerange{\pageref{firstpage}--\pageref{lastpage}} \pubyear{}

\maketitle

\label{firstpage}

\begin{abstract}
The second known outburst of the WZ Sge type dwarf nova GW Lib was
observed in April 2007. We have obtained unique multiwavelength data
of this outburst which lasted $\sim$ 26 days. {\it AAVSO} observers
recorded the outburst in the optical, which was also monitored by {\it
WASP}, with a peak V magnitude of $\sim$ 8. The outburst was followed
in the UV and X-ray wavelengths by the {\it Swift} UVOT and XRT
telescopes. The X-ray flux at optical maximum was found to be three
orders of magnitude above the pre-outburst quiescent level, whereas
X-rays are normally suppressed during dwarf nova outbursts. A distinct
supersoft X-ray component was also detected at optical maximum, which
probably arises from an optically-thick boundary layer.  Follow-up
{\it Swift} observations taken one and two years after the outburst
show that the post-outburst quiescent X-ray flux remains an order of
magnitude higher than the pre-outburst flux. The long interoutburst
timescale of GW Lib with no observed normal outbursts support the idea
that the inner disc in GW Lib is evacuated or the disc viscosity is
very low.
\end{abstract}

\begin{keywords}
accretion, accretion discs -- stars: dwarf novae -- stars: novae,
cataclysmic variables -- X-rays: stars -- X-rays: binaries
\end{keywords}

\section{Introduction}
Dwarf novae (DNe) are non-magnetic cataclysmic variables (CVs) with
accretion discs and with a white dwarf primary and a main sequence
secondary star. They are relatively nearby sources which provide a
laboratory for studying accretion disc physics in our Galaxy. The
orbital periods of DNe are typically between 70 min and 10 h.  From
time to time, the disc goes into an outburst which is a brightening of
the disc by 2--9 magnitudes and which can last from days to several
weeks. The mechanism leading to a dwarf nova outburst is thought to be
a disc instability which was first proposed by \citet{hos} (for a more
detailed discussion see \citet{las01}). After this luminous
phenomenon, the system returns to quiescence which can last from
$\sim$ 10 days to decades. The disc dominates the optical emission
during an outburst. SU UMa type dwarf novae show superoutbursts which
can last for several weeks and are characterized by superhumps,
periodic brightenings whose recurrence times are slightly longer than
the orbital period. Superhumps are thought to be due to a 3:1
resonance in the accretion disc \citep{whi88}. While most SU UMa types
show normal outbursts and superoutbursts, a subset called the WZ Sge
stars only have superoutbursts. A typical feature of superoutbursts is
a plateau phase in the optical lightcurve lasting for several days.

GW Lib was discovered in 1983 when it went into an outburst
\citep{maza}. It was present in ESO B Survey plates at magnitude 18.5
preceding the 1983 outburst. GW Lib brightened by 9 magnitudes during
the outburst and later faded back to the quiescent state and thus was
classified as a nova. Later studies showed that the spectrum resembled
a dwarf nova in quiescence \citep{due}. Since 1983, no other outbursts
of GW Lib had been observed until that of April 12, 2007
\citep{tem07}. This outburst, which was recorded by the {\it AAVSO}
observers, {\it WASP-South} and by {\it Swift}, lasted for 26
days. The brightest optical magnitude was reached at $\sim$ 8 mag in
the {\it V} band. GW Lib has been classified as a WZ Sge type star due
to its short period \citep[$P_{orb}$ = 76.78 min,][]{tho02} and low
accretion rate \citep{vanz04}. Typical characteristics of WZ Sge type
stars are short orbital periods, low mass-transfer rates and extremely
long recurrence times which can last for decades. GW Lib was the first
observed CV in which the accreting white dwarf showed non-radial
pulsations \citep{war98}. This phenomenon was not expected to be
discovered in accreting binaries since they were considered to be too
hot to be located in the DAV instability strip, although low mass
transfer rates from the secondary would explain low net accretion
rates onto the white dwarf, and thus a lower accretion heating of the
white dwarf \citep{vanz00}. The pulsations of white dwarfs are thought
to be due to g-mode non-radial gravity waves \citep{koe90}. In GW Lib,
pulsations are seen near 230, 370 and 650 s in the optical waveband
(e.g. {\it SAAO} data), and also in the {\it HST} UV data but with
$\sim$ 6 times higher amplitudes than in the optical \citep{szk}. An
{\it XMM-Newton} observation of GW Lib obtained in 2005 during its
quiescent state reveals that these pulsations are also present in the
{\it XMM-Newton} Optical Monitor (OM) data, but not seen in the X-ray
data \citep{hil}.  These {\it XMM-Newton} observations also confirmed
that GW~Lib has a very low accretion rate during quiescence.

In this paper, we present the 2007 outburst lightcurves of GW Lib in
the optical ({\it AAVSO} and {\it WASP-South}), UV ({\it Swift}, UVOT)
and X-ray ({\it Swift}, XRT) bands. We also present X-ray spectral
analysis of the outburst observations and follow-up {\it Swift}
observations of GW Lib in 2008 and 2009. Previous multiwavelength
observations covering outbursts of SU UMa (and WZ Sge) type systems
have been obtained from e.g. VW Hyi \citep{pri87,whe96b}, WZ Sge
\citep{kuu02,whe05}, and OY Car, which was observed by the {\it EUVE}
\citep{mau00}. Multiwavelength observations of dwarf novae are needed
in order to enhance our knowledge of astrophysical systems with discs,
e.g., X-ray binaries and AGNs. DN outbursts offer a good opportunity
to study the disc physics, and compare the current theory of outbursts
with the observational information. The fact that GW Lib does not show
any normal outbursts and the recurrence time between the two major
outbursts was over 20 years, suggests that the disc structure could be
different from most other dwarf novae.

\section{Observations and data reduction}
GW Lib was initially observed by the {\it Swift Gamma-ray Burst
Explorer} \citep{geh} between April 13th and May 16th, 2007, over an
interval of 30 days. The data were obtained with the
Ultraviolet/Optical Telescope (UVOT) \citep{rom} and with the X-ray
telescope (XRT) \citep{bur} which has an energy resolution of 140 eV
at 5.9 keV (at launch) (see \citet{cap05}, R $\sim$ 40). The XRT was
operating in the Window Timing (WT) and Photon-Counting (PC) modes
during the observations. The UVOT observations were obtained in the
imaging mode with the UV grism in order to provide spectral
information and to mitigate against coincidence losses. The resolution
of the UV grism is R $\sim$ 150 for 11--15 magnitude range
stars\footnote{http://heasarc.gsfc.nasa.gov/docs/swift/analysis/uvot\_ugrism.html}
The details of the 38 observations are listed in
Table~\ref{observations}. The first two observations were obtained in
imaging mode with the UVM2 filter, but these suffer from severe
coincidence loss effects and were thus excluded from our analysis.
The optical data were provided by the {\it AAVSO} observers and by the
{\it Wide Angle Search for Planets} (WASP) \citep{pol06}.

\begin{table*}
\centering
\caption{The {\it Swift} observations of GW Lib obtained in 2007, 2008 and 
2009.}
\begin{tabular}{rcccccrr}
\hline
\hline
 OBSID & T$_{start}$ & Roll angle &  UVOT & XRT WT & XRT PC & Ugrism \\
       &        & degrees    &  exposure (s) & exposure (s) & exposure (s) & exposure (s)\\
\hline								    
 30917001 & 2007-04-13T15:19:47 & 133 &  4914 & 2902 & 1910 & 0   \\ 
 30917002 & 2007-04-18T06:22:00 & 136 &  4781 & 4755 &    0 & 0   \\
 30917003 & 2007-04-20T17:53:57 & 141 &   992 &    5 &  990 & 992 \\
 30917004 & 2007-04-21T03:47:02 & 142 &   769 &    4 &  772 & 769 \\
 30917005 & 2007-04-21T14:55:01 & 142 &   904 &    2 &  910 & 904 \\
 30917006 & 2007-04-21T22:56:00 & 142 &  1442 &    4 & 1447 & 1442\\
 30917008 & 2007-04-22T23:02:00 & 142 &  1264 &    2 & 1271 & 1264\\
 30917009 & 2007-04-23T11:42:01 & 142 &  1050 &    8 & 1048 & 1050\\
 30917010 & 2007-04-24T00:50:00 & 142 &  1352 &   13 & 1349 & 1352\\
 30917011 & 2007-04-24T13:38:01 & 142 &  1288 &    4 & 1294 & 1288\\
 30917012 & 2007-04-25T02:38:00 & 142 &  1635 &    6 & 1639 & 1635\\
 30917013 & 2007-04-25T12:15:01 & 142 &  1632 &   10 & 1634 & 1632\\
 30917014 & 2007-04-26T02:50:01 & 142 &  1393 &    5 & 1400 & 1393\\
 30917015 & 2007-04-26T20:30:01 & 142 &  1344 &    6 & 1350 & 1344\\
 30917016 & 2007-04-27T02:28:00 & 155 &  5868 &   12 & 5890 & 5868\\
 30917017 & 2007-04-29T01:05:00 & 155 &  5643 &   15 & 5659 & 5643\\
 30917018 & 2007-05-01T01:18:01 & 168 &  3150 &    6 & 3162 & 3150\\
 30917019 & 2007-05-05T09:46:01 & 168 &  1597 &    3 & 1605 & 1597\\
 30917020 & 2007-05-05T22:38:01 & 168 &  1596 &    2 & 1605 & 1596\\
 30917022 & 2007-05-07T00:20:00 & 151 &  1589 &    9 & 1588 & 1589\\
 30917023 & 2007-05-07T11:35:01 & 151 &  1358 &    2 & 1364 & 1358\\
 30917024 & 2007-05-07T22:50:00 & 151 &  1595 &    7 & 1590 & 1595\\
 30917025 & 2007-05-08T10:09:00 & 156 &  1294 &    2 & 1294 & 1294\\
 30917026 & 2007-05-08T21:24:01 & 156 &  1247 &    2 & 1249 & 1247\\
 30917027 & 2007-05-09T08:41:00 & 156 &   875 &    6 &  872 &  875\\
 30917028 & 2007-05-09T19:49:01 & 156 &  1496 &    2 & 1504 & 1496\\
 30917029 & 2007-05-10T00:45:01 & 162 &  1457 &    4 & 1462 & 1457\\
 30917030 & 2007-05-10T18:19:00 & 162 &  1577 &    4 & 1582 & 1577\\
 30917031 & 2007-05-11T07:10:01 & 162 &  1433 &    3 & 1439 & 1433\\
 30917032 & 2007-05-11T18:25:01 & 162 &  1369 &    5 & 1371 & 1369\\
 30917033 & 2007-05-12T07:16:00 & 162 &  1372 &   10 & 1366 & 1372\\
 30917034 & 2007-05-12T18:31:00 & 162 &  1372 &    2 & 1379 & 1372\\
 30917035 & 2007-05-13T05:46:01 & 162 &  1157 &    4 & 1161 & 1157\\
 30917036 & 2007-05-13T18:37:00 & 162 &  1313 &    4 & 1319 & 1313\\
 30917037 & 2007-05-14T07:28:01 & 162 &  1373 &    4 & 1377 & 1373\\
 30917038 & 2007-05-14T18:42:01 & 162 &  1250 &    3 & 1255 & 1250\\
 30917039 & 2007-05-15T07:34:01 & 162 &  1400 &    2 & 1406 & 1400\\
 30917041 & 2007-05-16T04:27:01 & 162 &  1947 &    6 & 1956 & 1947\\
 30917042 & 2008-04-24T17:33:01 & 150 &  4772 &    0 & 4959 &    0\\
 30917043 & 2008-05-01T18:25:09 & 173 &    83 &    0 &  122 &    0\\
 30917044 & 2008-05-08T12:38:01 & 151 &  1612 &    0 & 1627 &    0\\                   
 30917045 & 2009-02-22T02:16:00 & 111 &  5939 &    0 & 5779 &    0\\
 30917046 & 2009-03-02T03:10:01 & 112 &  4536 &    0 & 4586 &    0\\
\hline
\hline
\label{observations}
\end{tabular}
\end{table*}

In addition to the outburst observations, follow-up {\it Swift} observations 
of GW~Lib were made in April-May 2008 and February-March 2009. For these 
observations the {\it V} and {\it UVW1} filters were used in the UVOT 
instrument. The details of these observations are also given in 
Table~\ref{observations}. 

\subsection{UV grism data reduction}
The {\it Swift} UV grism was operating in clocked mode at the time of
the outburst observations. We have used our own automated pipeline,
based around UVOT {\it Swift} FTOOLS \citep{imm} for the spectral
extraction of the UV grism data. The release version which we used
(\textsc{heasoft} v.6.1.2) of the {\it Swift} UVOT analysis software
does not allow grism spectra to be traced, nor extracted
optimally. Thus under FTOOLS, we were restricted to using a box
extraction of a fixed width for both source and background
regions. The extraction box must be well-centered, aligned in so far
as is possible with the dispersion direction and broad enough that
slight miscenterings do not preferentially exclude flux from the wings
of the extracted spectra at specific wavelengths. For a well-centered
box, a width of 35 pixels contains 90 per cent of the integrated source
counts, and this was the width chosen. Larger widths increase the
source counts, but can also add in poorly subtracted background
counts, and thus do not enhance the statistical quality, and further
increases the likelihood of contamination by nearby sources. For the
background, we used extraction widths of 25 pixels.

After having chosen our source and background regions for each
observation, the images were corrected for mod-8 fixed pattern noise,
and source and background spectra were extracted automatically using
the FTOOL \textsc{uvotimgrism}. Response functions were then created
with \textsc{uvotrmfgen}. Unfortunately, the last 39 spectra from 
10 May 2007 onwards were taken at a roll-angle which placed a nearby
source (TYC6766-1570-1) on a line between our source and in the same
direction as the dispersion direction. These spectra were severely
contaminated and we therefore exclude them from our analysis. The
wavelength range of the UV data for the lightcurve was restricted to
2200--4000 \AA\, due to contamination by other 0th order spectra in
the short wavelength end.

The 2008 and 2009 observations were obtained in filter mode. We derived
the {\it V} and {\it UVW1} ($\overline\lambda$ = 2600 \AA) fluxes and
magnitudes using the FTOOL \textsc{uvotmaghist}. The source counts
were extracted by using circular extraction region with r$_{src}$ = 5
arcsec. A circular background region with a radius of r$_{bg}$ = 10
arcsec was placed on a source free region in the field of view.

\subsection{X-ray data reduction}
The X-ray data reduction was performed with the standard \emph{Swift}
pipelines retaining grades 0--12 for the PC mode and 0--2 for the WT
mode data. The {\it Swift} X-ray data cover the energy range 0.3--10
keV, but the source is detected only up to 8.0 keV. The X-ray
lightcurve was extracted from the WT and PC mode data by using the
tools described in \citet{eva09}.

The X-ray spectra were extracted with \textsc{xselect}. For the WT
mode data, a 40 pixel (94'') box for the source and an 80 pixel
(189'') box for the background region were used. The normal limit for
pile-up in WT mode data is $\sim$ 100 ct/s \citep{romano}, thus it was
not necessary to account for this when extracting the spectra. The PC
mode source data were extracted using a 20 pixel (47'') circle and the
background data within a 60 pixel (141'') circular area. The PC mode
data from the first observation were not used for spectral
extraction. The PSFs of the rest of the PC mode data were checked to
verify that pile-up did not occur.

\begin{figure*}
\includegraphics[width=10cm]{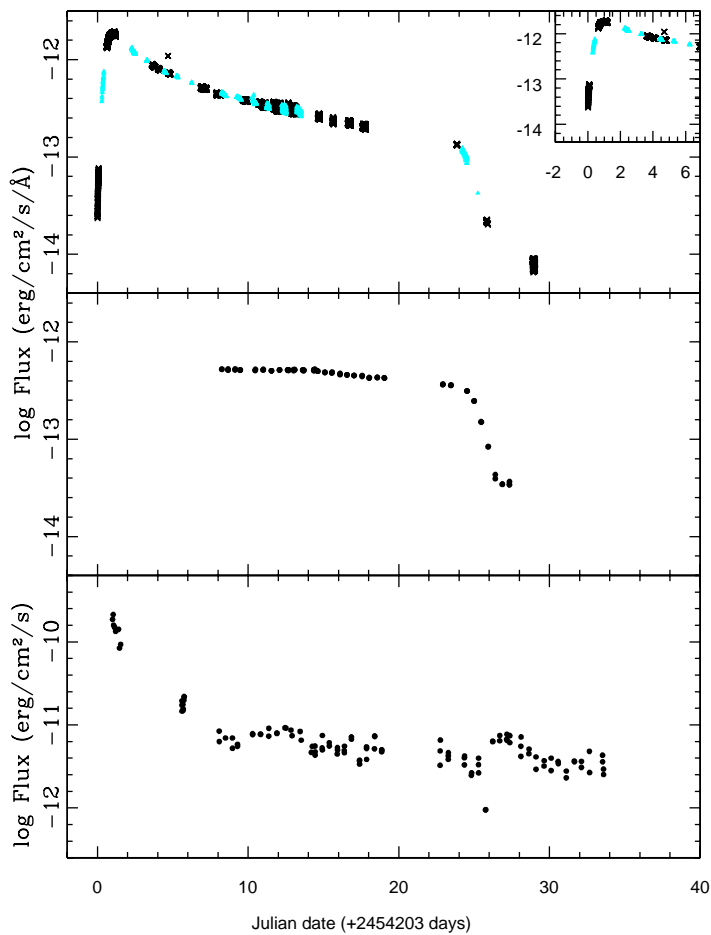}
\centering
\caption{The optical, UV and X-ray outburst lightcurves of GW
Lib. T$_{0}$ is JD 2454203.1, which corresponds to the start
time of the first data point in the {\it AAVSO} lightcurve. The top
panel shows the {\it AAVSO} optical lightcurve (www.aavso.org) in the
V band (converted to log F) in black and the {\it WASP} lightcurve
in light grey (see text for details). The upper panel insert shows the
rapid rise to maximum peaking one day after the onset of the
outburst. The middle panel shows the UV in the wavelength range
2200--4000 \AA\, (Swift UV grism), and the bottom panel the X-ray
lightcurve in 0.3--10.0 keV ({\it Swift} XRT). The X-ray data have been
binned at 600 s and the other bands are plotted per exposure.}
\label{gwliblc}
\end{figure*}

\subsection{WASP data reduction}
GW Lib falls into a field which was being intensively monitored by
the {\it WASP} search for transiting exoplanets \citep{pol06}
at the time of the outburst. The {\it WASP} is a wide-field
imaging system with each instrument having a field of view of 482
deg$^{2}$. The photometric accuracy of the instruments
is better than 1 per cent for objects with {\it V} magnitude $\sim$
7.0--11.5. The field (centred on 15h 02m, 28d 22m) was observed
between February 16, 2007, and July 19, 2007, with an average of 88
images per night, each with an exposure time of 30 s. The
data were extracted and calibrated in automated fashion using the
standard {\it WASP} pipeline, full details of which are given in
\citet{pol06}. 

\section{Time series analysis}

\subsection{Outburst lightcurves}
Fig.~\ref{gwliblc} presents the April-May 2007 outburst of GW Lib. The
top panel shows the {\it AAVSO} V-band lightcurve in black and the
{\it WASP} lightcurve ({\it V}+{\it R} bands) in light grey (scaled to
match the {\it AAVSO} {\it V} band magnitude). The magnitudes have
been converted to log F in order to make comparison between the
optical and UV lightcurves clearer. The middle and lower panels show
the UV and X-ray lightcurves respectively. The {\it AAVSO} outburst
lightcurve is already rising from magnitude 13 (log F $\approx$ -13.6
erg cm$^{-2}$ s$^{-1}$ \AA$^{-1}$) on JD 2454203.1 which we define as
T$_{0}$. It reaches its highest peak one day later at magnitude 7.9
(log F $\approx$ -11.7 erg cm$^{-2}$ s$^{-1}$ \AA$^{-1}$), and after
the peak, declines steadily for 23 days. At 29 days after the
beginning of the outburst the optical brightness declines sharply to
magnitude $\sim$ 14 (log F $\approx$ -14.2 erg cm$^{-2}$ s$^{-1}$
\AA$^{-1}$), presumably as the accretion disc returns to its quiescent
state.  Large variations in the {\it WASP} lightcurve around day
T$_{0}$ + 11, resulting from poor weather conditions were excluded.

The rise of the UV emission for GW Lib was not observed, and so we are
not able to quantify any UV delay, such as that previously measured by
e.g \citet{has83} in VW Hyi. However, the system is already extremely
UV bright at the time of the first {\it Swift} observations, 1 d after
the beginning of the outburst, when the UVM2 filter was used.
Unfortunately, these two observations are far too over-exposed to
provide useful flux measurements and they are not shown in our
lightcurve. Our UV lightcurve (from the UVOT grism data) starts 8 d
after the onset of the outburst. It remains almost flat for $\sim$ 6
days and then starts to decline. At 24.5 days after the optical rise
the UV lightcurve shows a steep decline, approximately simultaneously
with the sharp optical decline. It reaches a minimum about 2 d later.
The UV data after the day T$_{0}$ + 27.5 have been excluded due to
contamination.

The X-ray observations (0.3--10.0 keV) start 1 d after the rise in the
optical, approximately at the time of the optical peak. Our data do
not cover the rise of the X-rays, thus we are not able to quantify any
delay between the optical and X-ray rise, such as that observed e.g.\
by \citet{whea} in SS~Cyg. In GW Lib, the X-ray flux is initially
$\sim3$ orders of magnitude above its pre-outburst quiescent level
\citep{hil}, and it declines rapidly for $\sim$ 10 days before
continuing to decline more slowly during the remainder of the optical
outburst. Around day T$_{0}$ + 26, the flux dips sharply from log F
$\approx$ -11.6 erg cm$^{-2}$ s$^{-1}$ to log F $\approx$ -12.0 erg
cm$^{-2}$ s$^{-1}$, lasting for less than a day. We have checked these
data carefully and it is clear that this is a real dip in the
brightness of the source.  After the dip, the X-rays rise to the level
of log F $\approx$ -11.2 erg cm$^{-2}$ s$^{-1}$ and show a bump
between days T$_{0}$ + 26 and T$_{0}$ + 29, which peaks at
approximately the same time that the UV lightcurve finishes its steep
decline. From T$_{0}$ + 29 days onwards, the X-ray flux remains
approximately constant at $\sim$ -11.5 erg cm$^{-2}$ s$^{-1}$.

During the two follow-up {\it Swift} observations in 2008 the count
rate was consistent on days T$_{0}$ + 378 and T$_{0}$ + 392 at 0.03
$\pm$ 0.01 ct s$^{-1}$. In 2009, 682 and 690 days since T$_{0}$, it
was 0.04 $\pm$ 0.01 ct s$^{-1}$.  For comparison, \citet{hil} report a
quiescence count rate of 0.02 ct s$^{-1}$ (0.2--10.0 keV) in the {\it
XMM-Newton} pn camera on August 25--26, 2005, about two years before
the outburst. Assuming a thermal bremsstrahlung model of kT = 3 keV,
this corresponds to an XRT count rate of 0.002 ct s$^{-1}$, where we
used
WebPIMMS\footnote{http://heasarc.gsfc.nasa.gov/Tools/w3pimms.html} to
make the conversion. Thus the X-ray flux seems to have remained an
order of magnitude above the pre-outburst quiescent level for about
two years after the outburst.

\begin{figure}
\centering
\includegraphics[width=8cm]{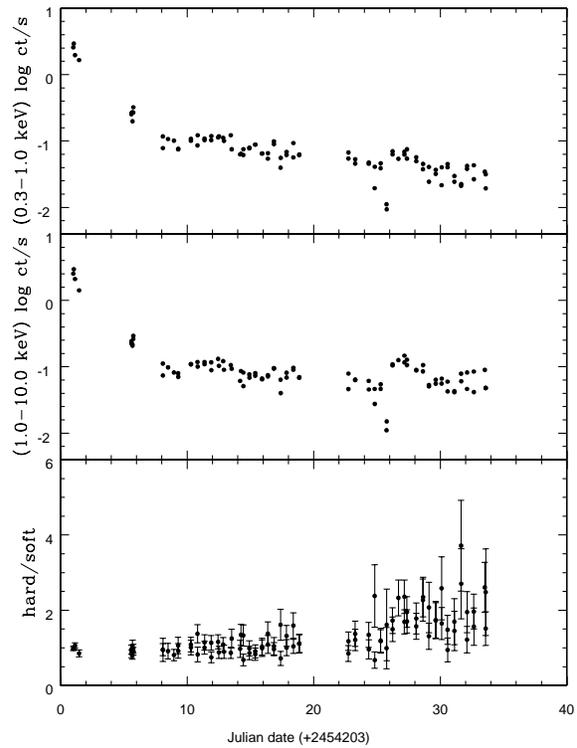}
\caption{The soft (0.3--1.0 keV, upper panel) lightcurve, the hard
(1.0--10.0 keV, middle panel) lightcurve, and the X-ray hardness ratio
(hard/soft) of GW Lib. The data have been binned at 700 s.}
\label{hardness}
\end{figure}

Fig.~\ref{hardness} shows the soft X-ray lightcurve (upper panel) in
the 0.3--1.0 keV band and the hard X-ray lightcurve (middle panel) in
the 1.0--10.0 keV band from the outburst observations. The lower panel
shows the hardness ratio \citep[extracted by using the tools
of][]{eva09}. The hardness ratio gradually increases during the
outburst, and shows a sharp increase at the time of the bump in the
X-ray light curve, which coincides with the steep decline in the
optical and UV lightcurves.

\subsection{Search for oscillations in the X-ray data}
Pulsations were discovered in the quiescent optical emission of GW~Lib
by \citet{war98}. Pulsation periods near 230, 370 and 650 s are seen
in the optical and UV wavebands \citep[e.g.][]{vanz04,szk} and an
optical 2.1 h modulation was discovered by \citet{wou02}.
\citet{cop09} found that the pulsation periods were suppressed after
the 2007 outburst, but that the 2.1 h modulation remained.

\citet{hil} searched for these modulations in the {\it XMM-Newton} X-ray
data taken during quiescence before the outburst, but did not detect
any significant periodicities. They measured an upper limit of 0.092
mag for the X-ray pulsations.

We searched for periodicities from the outburst data by using two
methods. When examining shorter timescales we took power spectra of
near continuous sections of data, applying the normalisation of
\citet{lea83} which allows the noise powers to be easily characterised
so that detection limits can be set \citep[see also][]{lew88}. For
this method, the WT mode data were binned into 1 s time bins and the
PC mode data into 5 s bins, which gave approximately 512 and 256 time
bins per Fourier transform from individual {\it Swift} orbits,
respectively. This gave 13 transforms for the WT mode and 48 for the
PC mode, from which the averaged power spectrum was constructed for
each mode. At the 99 per cent confidence level, no periodicities were
seen in the data with a fractional amplitude upper limit of 6 per cent
(WT mode) over the period range 2--512 s, and 11 per cent (PC mode)
over the period range 10--640 s.

When searching for longer timescale modulations, notably at 650 s and
2.1 hr ($\sim$ 7560 s), as seen previously by the authors mentioned
above, we initially calculated Lomb-Scargle periodograms of the
data. Unfortunately, the identification of any potentially real
modulation in the WT mode data at these timescales was hindered by the
window function caused by the light curve sampling. However, by
folding the PC mode data at the previously seen longer periods and at
the orbital period \citep[P$_{orb}$ = 76.78 min,][]{tho02} we
estimated 99 per cent upper limits to the fractional amplitude of any
modulation at 650 s, 76.78 min ($\sim$ 4607 s) and 7560 s to be 7, 6
and 8 per cent, respectively.

\section{Spectral analysis}

\subsection{Outburst X-ray spectra}

\begin{figure*}
\includegraphics[width=90mm,angle=-90]{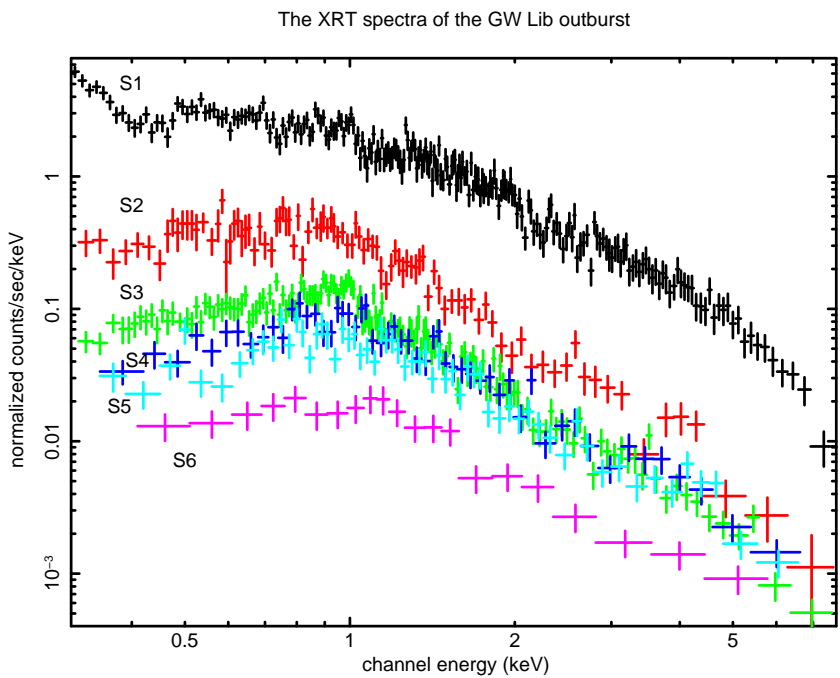}
\centering
\caption{The X-ray spectra of GW Lib throughout the outburst.}
\label{xrayoutburst}
\end{figure*}

Five X-ray spectra were extracted from the outburst observations to
investigate the spectral evolution. The first and the second spectra,
S1 and S2, consist of single observations (WT mode data) covering days
T$_{0}$ + 1--2 and 6 respectively. The last three spectra, S3, S4 and
S5, were extracted by combining observations covering longer time
intervals (PC mode data). These intervals are given in
Table~\ref{spectra}. The X-ray spectra were binned at 20 cts/bin
using the FTOOL \emph{grppha}, and the spectral analysis was carried
out using Xspec11. Fig.~\ref{xrayoutburst} shows the X-ray spectra
from the start of the outburst (top) until the end (bottom). The
combined spectrum from the follow-up observations of 2008 and 2009 is
also shown (labeled S6). The outburst spectra show that the increase
in hardness apparent in Fig.~\ref{hardness} is due to a decrease in
the relative strength of emission below about 1\,keV. There is also
excess soft X-ray emission apparent below 0.4\,keV in the first
observation (S1).

\begin{table}
\caption{The XRT spectra of GW Lib and the corresponding
observations. The epoch is in days since T$_{0}$.}
\centering
\begin{tabular}{cc}
\hline
\hline       
Spec &  Epoch  \\
     &   (d)   \\
\hline
 S1  & 1--2    \\ 
 S2  &   6     \\
 S3  & 8--19   \\
 S4  & 23--28  \\
 S5  & 28--34  \\
 S6  & (2008 \& 2009 data)  \\
\hline
\hline
\label{spectra}
\end{tabular}
\end{table}

When DNe are in quiescence, the boundary layer is optically-thin and
assumed to be a source of hard X-rays \citep{pri79}. During an
outburst, the boundary layer is optically-thick and emits more
luminous soft X-ray emission which is usually described as black body
emission \citep{pri77}. The optically-thin hard X-ray component is
usually somewhat suppressed in outburst.

\begin{table*}
\caption{The results of the spectral fitting using photoelectric
absorption, black body and three optically-thin emission components.
Not all components are fitted to all spectra.  The errors correspond
to the 90 per cent confidence limits for one parameter of
interest. The measured abundance from the S1 spectrum is
0.02$^{+0.01}_{-0.01}$, and the abundances in the other fits have been
fixed at this value. The unabsorbed fluxes F$_{1}$, F$_{2}$ and
F$_{3}$ correspond to the three optically-thin emission components in
the 0.3--10.0 keV band.}  \centering
\begin{tabular}{cccccc}
\\
\hline
\hline
Spectrum & S1 & S2 & S3 & S4 & S5 \\
\\
Model & wa(bb+3me) & wa(3me) & wa(2me) & wa(2me) &  wa(2me) \\
\hline 
n$_{H}$ & 23.83$^{+0.05}_{-0.04}$ & 8.61$^{+5.38}_{-3.99}$ & 8.15$^{+1.23}_{-1.17}$ & 13.25$^{+7.28}_{-3.38}$ & 14.63$^{+6.86}_{-4.36}$ \\
(10$^{20}$ cm$^{-2}$) & & & & \\
kT$_{bb}$ & 0.013$^{+0.001}_{-0.001}$ & -- & -- & -- & --   \\
(keV)
\\
\\
kT$_{1}$ & 5.46$^{+1.26}_{-0.86}$ & 4.80$^{+6.07}_{-1.66}$  & 5.79$^{+2.73}_{-1.51}$ & 5.19$^{+5.13}_{-1.88}$ &  4.92$^{+3.82}_{-1.36}$ \\
(keV)
\\
\\
kT$_{2}$ & 0.71$^{+0.23}_{-0.13}$ & 0.64$^{+0.21}_{-0.14}$ & 0.66$^{+0.06}_{-0.07}$ & 0.57$^{+0.23}_{-0.24}$ &  0.48$^{+0.20}_{-0.16}$  \\
(keV)
\\
\\
kT$_{3}$ & 0.17$^{+0.01}_{-0.01}$ & 0.17$^{+0.10}_{-0.05}$ & -- & -- & -- \\
(keV)
\\
\\
F$_{1}$ & 120. & 6.72 & 3.05 & 3.88 & 3.62 \\
($10^{-12}$ erg/cm$^{2}$/s)
\\
\\
F$_{2}$ & 30.5 & 8.44 & 3.94 & 2.38 & 1.67 \\
($10^{-12}$ erg/cm$^{2}$/s)
\\
\\
F$_{3}$ & 271. & 4.00 & -- & -- & -- \\
($10^{-12}$ erg/cm$^{2}$/s)
\\
\\
$\chi^{2}_{\nu}$/$\nu$ & 1.08/220 & 1.25/89 & 1.13/150 & 0.94/50 & 1.19/56  \\
\hline
\hline
\label{fitnumbers2}
\end{tabular}
\end{table*}

The {\it Swift} XRT spectra shown in Fig.~\ref{xrayoutburst} are
clearly dominated by hard X-ray emission, and so we started by fitting
the first spectrum (S1) using an optically-thin thermal plasma model
\citep[\textsc{mekal} model in Xspec,][]{mew86,lie95} with
photoelectric absorption \citep[\textsc{wabs},][]{mor83} letting the
abundance vary freely. This fit was not successful
($\chi^{2}_{\nu}$/$\nu$ = 3.35/226) and a clear excess of soft photons
was present below 0.7 keV. To account for this excess we then added a
black body component to the fit which improved ($\chi^{2}_{\nu}$/$\nu$
= 2.54/224). There were still strong residuals between 0.5--0.7 keV
suggesting that a second optically-thin thermal emission model was
needed.  The abundances were still allowed to vary but they were tied
between the two optically-thin emission models. With this model (a
black body and two optically-thin emission components with
photoelectric absorption), we obtained a statistically acceptable fit
to the S1 spectrum ($\chi^{2}_{\nu}$/$\nu$ = 1.11/222). We tested
whether a third optically-thin emission component would improve the
fit, and found a slightly better fit statistic of
$\chi^{2}_{\nu}$/$\nu$ = 1.08/220. Using the F-test, we found that
this provides a better description of the underlying spectrum with a
confidence of 98.5 per cent. Of course, these three distinct
temperature components are most likely just an approximation to an
underlying continuous distribution, such as the cooling flow models
used by e.g.\ \citet{whe96b} and \citet{muk}.

\begin{figure*}
\includegraphics[width=84mm,angle=-90]{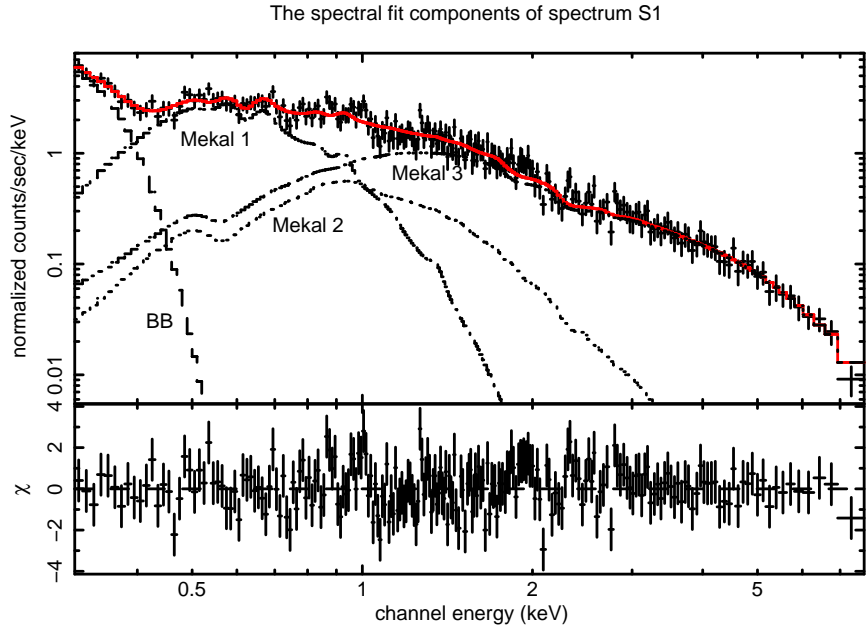}
\centering
\caption{Upper panel: best-fit model components of the first outburst
spectrum (S1) of GW Lib. The data are fitted with a black body and
three optically-thin thermal emission components, all absorbed by the
same column density. Lower panel shows the residuals.}
\label{SPEC1}
\end{figure*}

\begin{figure*}
\includegraphics[width=84mm,angle=-90]{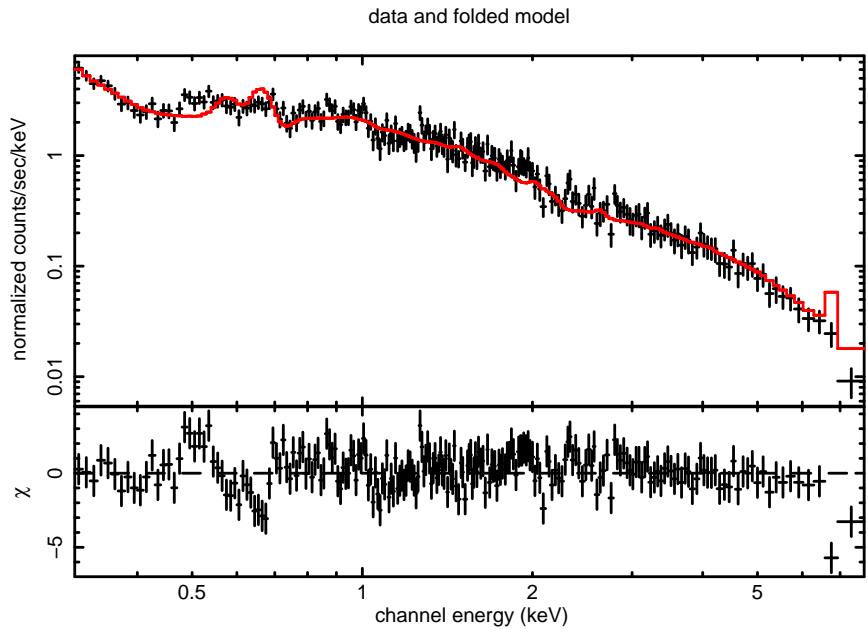}
\centering
\caption{Upper panel: a spectral fit (a black body and three
optically-thin thermal emission models with variable abundances
(\textsc{vmekal})) to the first outburst spectrum (S1) with the oxygen
abundance fixed to that found in the {\it XMM-Newton} quiescent
spectrum by \citet{hil}. An oxygen line of this strength (at
0.65\,keV) would be detected and the fit is unacceptable with
$\chi^{2}_{\nu}$/$\nu$ = 1.62/221. The lower panel shows clear
residuals between 0.55--0.70 and above 6.5 keV.}
\label{fixed_oxygen}
\end{figure*}

The data and the best-fit model components for S1 with residuals are
plotted in Fig.~\ref{SPEC1}. The full set of fitted parameters is
given in Table~\ref{fitnumbers2}. It is worth noting that the fitted
abundance is low, only 0.02$^{+0.01}_{-0.01} \times$ solar. We looked
at this in some detail and found that the low abundance in our fit is
driven entirely by the lack of thermal iron lines at around
6.7\,keV. Strong line emission is predicted also by the model around
1\,keV, but the spectral resolution here is not sufficient to provide
strong constraints on abundances. Nevertheless, by fitting with a
black body and three optically-thin thermal emission models with
variable abundances (\textsc{vmekal}), we found that the high oxygen
abundance reported by \citet{hil} in GW Lib in quiescence (6--8
$\times$ solar) can be ruled out for our S1 spectrum. Setting the
oxygen abundance to 6 $\times$ solar in our model resulted in an
unacceptable best-fit statistic of $\chi^{2}_{\nu}$/$\nu$ =
1.62/221. This fit is plotted in Fig.~\ref{fixed_oxygen}.

In order to investigate the evolution of the X-ray spectra of GW Lib
during the outburst, we tested the need for multi-temperature
optically-thin emission components and for the black body in a similar
manner for the second spectrum S2 and the subsequent spectra S3, S4
and S5. Since the abundances of the spectra S2, S3, S4 and S5 were not
well-constrained, they were fixed to the best fit abundance of
spectrum S1. There is no evidence for the presence of a black body
component in these spectra. Thus, a black body was not employed in the
spectral fitting of the subsequent outburst spectra. The best fit for
spectrum S2 was found with a photoelectric absorption and three
optically-thin emission components which yielded a goodness of fit of
$\chi^{2}_{\nu}$/$\nu$ = 1.25/89. The rest of the outburst spectra S3,
S4 and S5 were fitted successfully with photoelectric absorption and
two optically-thin emission components. The best fit values, their 90
per cent confidence limits and the unabsorbed flux for each thermal
emission component in the 0.3--10.0 keV range are given in
Table~\ref{fitnumbers2}. It can be seen that the increase in hardness
towards the end of the outburst is due to a decrease in the relative
quantity of cooler gas as the flux of the cooler temperature
component drops in spectra S3-S5.

\subsection{Fluxes, luminosities and accretion rates}
\begin{figure*}
\includegraphics[width=90mm,angle=-90]{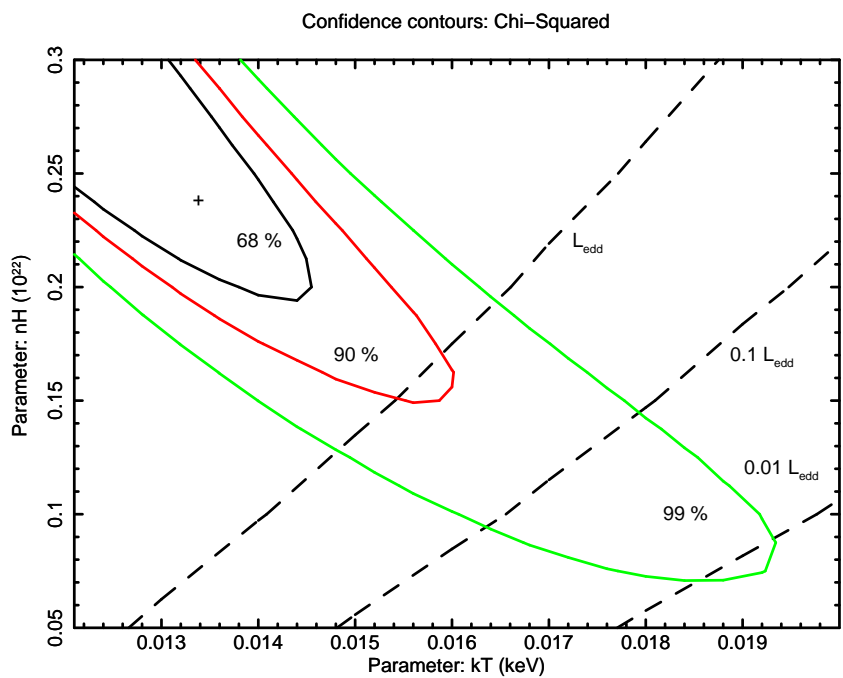}
\centering
\caption{Black body parameters derived from the first outburst
spectrum (S1) of GW Lib. The 10$^{38}$ (L$_{edd}$), 10$^{37}$ (0.1
L$_{edd}$) and 10$^{36}$ erg s$^{-1}$ (0.01 L$_{edd}$) bolometric
luminosity levels of the black body component are shown as dashed
lines. The contours describe the 68, 90 and 99 per cent confidence
levels for n$_{H}$ v. kT$_{bb}$.}
\label{bolom}
\end{figure*}

The total fluxes, luminosities and accretion rates in the 0.3--10 keV
range are given in Table~\ref{luminosity}. The fluxes given in the
0.3--10 keV range are absorbed fluxes. The distance was taken to be r
= 104\,pc \citep{tho03}. The accretion rates were estimated by using
$\dot{M}$ = $2LR_{WD}$/$GM_{WD}$, where we adopt M$_{WD}$ = 1
M$_{\odot}$ and R$_{WD}$ = 5.5 $\times$ 10$^{8}$ cm \citep{tow04}.
The first X-ray spectrum gives a luminosity of 2.06 $\times$ 10$^{32}$
erg s$^{-1}$ which corresponds to $\dot{M}$ = 1.71 $\times$ 10$^{15}$
g s$^{-1}$ or 2.7 $\times$ 10$^{-11}$ M$_{\odot}$ yr$^{-1}$ in the
0.3--10 keV band. Since we do not see a rise in the X-rays, we are not
able to say whether this is the peak luminosity of this outburst. We
also derived the corresponding parameters for the bolometric
luminosity by extrapolating our spectral fit over the range
0.0001--100 keV.  The bolometric luminosity of the black body
component in spectrum S1 is not well constrained, and the best fit
value exceeds the Eddington limit of 1.3 $\times$ 10$^{38}$ erg
s$^{-1}$ for the assumed mass; Fig.~\ref{bolom} shows the 68, 90 and
99 per cent confidence levels of n$_{H}$ v. kT$_{bb}$ and the
corresponding range of bolometric luminosities. It can be seen that
fits are allowed with much lower absorption column densities and
luminosities. There is no reason to believe that the luminosity would
exceed the Eddington limit. Estimating the luminosities of supersoft
X-ray sources with black body models is notoriously unreliable
\citep[e.g.][]{kra96}. Consequently we do not attempt to give a
bolometric luminosity for S1 in Table~\ref{luminosity}.

\begin{table*}
\centering
\caption{The fluxes, luminosities and accretion rates in the 0.3--10
keV (fluxes absorbed) and 0.0001--100 keV (bolometric, fluxes
unabsorbed) bands of the XRT spectra.}
\begin{tabular}{ccccccc}
\\
\hline
\hline
Spec &    Flux         & Luminosity    &$\dot{M}$         & Flux          &Luminosity&$\dot{M}$ \\
     & erg cm$^{-2}$ s$^{-1}$  & erg s$^{-1}$  & g s$^{-1}$ & erg cm$^{-2}$ s$^{-1}$ & erg s$^{-1}$ & g s$^{-1}$  \\
     &(0.3--10 keV)  &(0.3--10 keV)&(0.3--10 keV) &(0.0001--100 keV)&(0.0001--100 keV)&(0.0001--100 keV)\\
\hline
S1   &1.60$\times$10$^{-10}$ & 2.06$\times$10$^{32}$ & 1.71$\times$10$^{15}$ & -- & --  & -- \\
S2   &1.25$\times$10$^{-11}$ & 1.63$\times$10$^{31}$ & 1.35$\times$10$^{14}$ &5.10$\times$10$^{-11}$&6.66$\times$10$^{31}$ & 5.52$\times$10$^{14}$\\
S3   &5.18$\times$10$^{-12}$ & 6.66$\times$10$^{30}$ & 5.52$\times$10$^{13}$ &1.12$\times$10$^{-11}$&1.44$\times$10$^{31}$ & 1.19$\times$10$^{14}$\\ 
S4   &4.38$\times$10$^{-12}$ & 5.63$\times$10$^{30}$ & 4.67$\times$10$^{13}$ &9.53$\times$10$^{-12}$&1.23$\times$10$^{31}$ & 1.02$\times$10$^{14}$\\
S5   &3.66$\times$10$^{-12}$ & 4.71$\times$10$^{30}$ & 3.89$\times$10$^{13}$ &8.12$\times$10$^{-12}$&1.04$\times$10$^{31}$ & 8.62$\times$10$^{13}$\\
\hline
\hline
\label{luminosity}
\end{tabular}
\end{table*}

\subsection{Later X-ray observations}
We also determined the flux and luminosity for the individual
integrated spectra of the 2008 and 2009 observations, and for the
combined 2008+2009 spectrum. The spectra were fitted with an absorbed
single-temperature optically-thin thermal emission model. For the
individual 2008 and 2009 spectra, we used Cash statistics due to the
low number of counts. The spectral fitting parameters with fluxes and
luminosities are given in Table~\ref{newparam}. The abundance was
unconstrained by the data and we fixed the value of Z = 0.02
$Z_{\odot}$ as obtained in outburst. The values in
Table~\ref{newparam} show that the state of GW Lib did not change much
between the 2008 and 2009 observations. Compared to the luminosity
obtained by \citet{hil} before the outburst (L(0.2--10 keV) = 9
$\times$ 10$^{28}$ erg s$^{-1}$), GW Lib is still an order of
magnitude brighter in the {\it Swift} observations $\sim$ two years
after the outburst.

\begin{table}
\caption{The spectral fitting parameters with fluxes and luminosities
of the 2008 and 2009 (columns 2 and 3) X-ray observations of GW
Lib. The fourth column shows the spectral fitting parameters for the
combined 2008 and 2009 spectrum.}  
\centering
\begin{tabular}{ccccc}
\hline
\hline       
Parameter & 2008 & 2009 & 2008+2009\\
\hline
n$_{H}$ & 7.23$^{+5.74}_{-5.08}$ & 5.63$^{+5.09}_{-4.32}$ & 0.11$^{+4.45}_{-0.11}$ \\
10$^{20}$ cm$^{-2}$ & & & \\ 
\\
kT & 1.20$^{+0.64}_{-0.33}$ & 2.00$^{+1.09}_{-0.64}$ & 3.51$^{+1.46}_{-1.16}$\\
keV & & & \\
\\
F(0.3--10 keV)&7.5$\times$10$^{-13}$ & 8.5$\times$10$^{-13}$ & 1.1$\times$10$^{-12}$\\
erg cm$^{-2}$ s$^{-1}$ & & \\
\\
L(0.3--10 keV) &9.8$\times$10$^{29}$ &1.1$\times$10$^{30}$ & 1.4$\times$10$^{30}$\\
erg s$^{-1}$ & & \\
\\
$\dot{M}$(0.3--10 keV)&8.2$\times$10$^{12}$ &9.1$\times$10$^{12}$ & 1.2$\times$10$^{13}$ \\
g s$^{-1}$ & & \\
\\
F(0.0001--100 keV) & 1.4$\times$10$^{-12}$ &1.4$\times$10$^{-12}$ & 1.3$\times$10$^{-12}$ \\
erg cm$^{-2}$ s$^{-1}$ & & \\
\\
L(0.0001--100 keV) &1.8$\times$10$^{30}$ &1.8$\times$10$^{30}$ & 1.7$\times$10$^{30}$ \\
erg s$^{-1}$ & & \\
\\
$\dot{M}$(0.0001--100 keV)&1.5$\times$10$^{13}$ &1.5$\times$10$^{13}$ & 1.4$\times$10$^{13}$ \\
g s$^{-1}$ & & \\
\hline
\hline
\label{newparam}
\end{tabular}
\end{table}

\subsection{Outburst UV spectra}
\begin{figure*}
\includegraphics[width=84mm,angle=0]{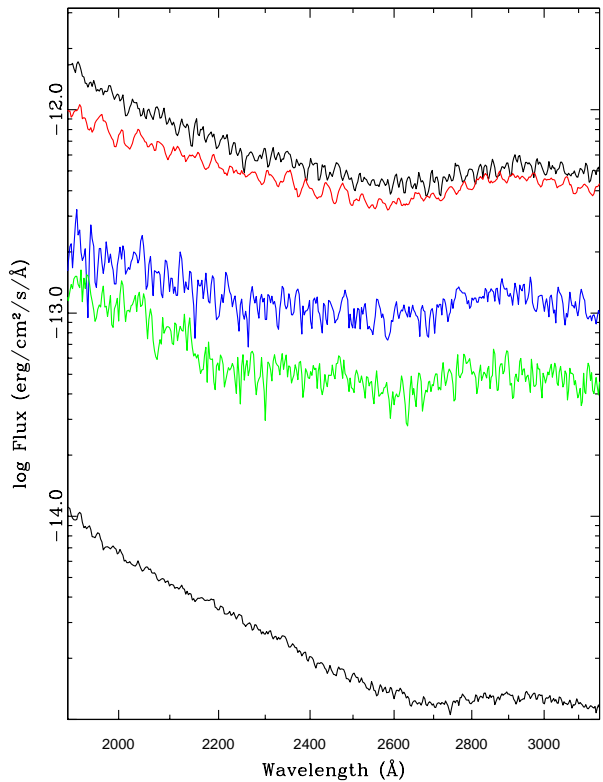}
\centering
\caption{The evolution of the background-subtracted UVOT UV grism flux
spectra of GW Lib throughout the outburst in descending and
chronological order. The top spectrum is an average of three spectra
corresponding to days T$_{0}$ + 8, 13 and 18. The following three
spectra correspond to days T$_{0}$ + 23, 26, and 27 (the spectrum on
day T$_{0}$ + 27 is an average of three spectra during that day). The
bottom spectrum shows the background flux level for one of the spectra
on day T$_{0}$ + 27.}
\label{uvspectra}
\end{figure*}

Four background-subtracted source UV outburst spectra and one
background spectrum are shown in Fig.~\ref{uvspectra}. The observing
dates corresponding to each spectrum are given in
Table~\ref{uv_obs_dates}. They were chosen to represent the evolution
of the UV data with time. The spectra show a blue continuum without
strong emission or absorption lines, although weak features would be
difficult to identify as the spectra suffer from modulo-8 fixed
pattern noise. The top spectrum is an average of three source spectra
corresponding to days T$_{0}$ + 8, 13 and 18 during the initial slow
decline in Fig.~\ref{gwliblc}. The two subsequent spectra are from
days T$_{0}$ + 23 (before the steep decline) and T$_{0}$ + 26 (after
the steep decline). The penultimate spectrum averages three spectra
around day T$_{0}$ + 27 and the bottom spectrum shows the background
flux for one of the source spectra on day T$_{0}$ + 27. It seems the
spectra become less blue after the sharp decline. All of the spectra,
including the background spectrum, show a bump around $\sim$ 2900 \AA.
This feature is most likely due to the contribution of the second
order spectrum (see ``Notes for observing with the UVOT UV
grism''\footnote{http://swift.gsfc.nasa.gov/docs/swift/analysis/uvot\_ugrism.html}
which shows that for a calibration white dwarf the second-order UV
light starts $\sim$ 2800 \AA). In the last two source spectra, the
background noise features are becoming clearer when the source itself
is becoming fainter.

The July 2001 {\it HST}/STIS outburst spectrum of WZ
Sge showed a Mg II absorption line at 2790--2810 \AA\,
\citep{kuu02}. \citet{kuu02} note that most of it originates from WZ
Sge and part of it is due to interstellar absorption. This line is not
seen in the UV spectra of GW Lib probably due to the much lower
resolution of the UV grism.

\begin{table}
\centering
\caption{The observation dates since the onset of the outburst for the
UV grism spectra in Fig.~\ref{uvspectra}.}
\label{uv_obs_dates}
\begin{tabular}{ccc}
\hline
\hline
T$_{0}$ + &  ObsDate   \\
\hline
 8       & 2007-04-20  \\  
13       & 2007-04-25  \\
18       & 2007-04-30  \\ 
23       & 2007-05-05  \\ 
26       & 2007-05-08  \\
27       & 2007-05-09  \\
\hline
\hline
\end{tabular}
\end{table}

The calibration of the UV grism is on-going, and consequently absolute
flux measurements remain uncertain. Nevertheless, we measured
monochromatic fluxes at 2200 \AA\, for the source spectra in
Fig.~\ref{uvspectra} and obtained $\lambda$F$_{2200}$ $\sim$ 1.49
$\times$ 10$^{-9}$, 1.14 $\times$ 10$^{-9}$, 2.13 $\times$ 10$^{-10}$,
and 1.15 $\times$ 10$^{-10}$ erg cm$^{-2}$ s$^{-1}$ respectively. For
$L$ = 4$\pi$r$^{2}$$F$ with r = 104 pc \citep{tho03}, the fluxes given
above correspond to monochromatic luminosities $\lambda$L$_{2200}$ of
$\sim$ 1.95 $\times$ 10$^{33}$, 1.49 $\times$ 10$^{33}$, 2.78 $\times$
10$^{32}$, and 1.50 $\times$ 10$^{32}$ erg s$^{-1}$.

\subsection{Later UV observations}
We studied the follow-up UVOT observations from 2008 and 2009 in order
to see how the UV magnitudes and fluxes have changed since the
outburst. The results are given in Table \ref{uvmagnitude}. Since the
last measurement in the outburst lightcurve, the UV flux has faded by
0.8 in log F.

\begin{table}
\caption{The average magnitudes, fluxes and their 1$\sigma$ errors in
the UVOT {\it V} and {\it UVW1} ($\overline\lambda$ = 2600 \AA)
filters for the 2008 and 2009 observations of GW Lib. The magnitudes
and fluxes are the mean values of different snapshots. Observation
time corresponds to the midpoint of each observation in days since
T$_{0}$. The exposure times are given in Table~\ref{observations}.}
\label{uvmagnitude}
\begin{tabular}{ccccc}
\hline
\hline
ObsID & Obstime & Filter  & Magnitude &     log Flux        \\
      &  d      &         &   mag     & erg s$^{-1}$ cm$^{-2}$ \AA$^{-1}$  \\
\hline
7042  & 378.25  & V    & 16.37 $\pm$ 0.01 & -14.979 $\pm$ 0.005 \\  
7042  & 378.25  & UVW1 & 14.69 $\pm$ 0.01 & -14.277 $\pm$ 0.002 \\
7043  & 385.21  & V    & 16.40 $\pm$ 0.07 & -14.987 $\pm$ 0.030 \\ 
7044  & 392.04  & V    & 16.39 $\pm$ 0.02 & -14.987 $\pm$ 0.009 \\
7044  & 392.04  & UVW1 & 14.60 $\pm$ 0.01 & -14.243 $\pm$ 0.004 \\
7045  & 682.08  & V    & 16.54 $\pm$ 0.01 & -15.042 $\pm$ 0.005 \\
7045  & 682.08  & UVW1 & 14.84 $\pm$ 0.01 & -14.339 $\pm$ 0.002 \\
7046  & 689.81  & V    & 16.53 $\pm$ 0.02 & -15.039 $\pm$ 0.006 \\
7046  & 689.81  & UVW1 & 14.84 $\pm$ 0.01 & -14.338 $\pm$ 0.003 \\
\hline
\hline
\end{tabular}
\end{table}
 
\section{Discussion}
In most cases, the hard X-ray emission of dwarf novae is suppressed
during an outburst. This is the case for VW~Hyi \citep{whe96b}, SS~Cyg
\citep{ric79,jon92,whea}, Z~Cam \citep{whe96a}, YZ Cnc \citep{ver99}
and WZ~Sge \citep{whe05}. In contrast, the outburst X-ray emission of
GW~Lib peaks at 2--3 orders of magnitude higher than its quiescent
level obtained by {\it XMM-Newton} in 2005 \citep{hil}. U~Gem is the
only other dwarf nova seen to increase its X-ray luminosity during an
outburst \citep{swa78}, and in this case the outburst emission is only
about a factor of five above its quiescent level \citep{mat00}.

The absolute luminosity of the peak X-ray emission of GW~Lib is high,
but not extraordinary (Table~\ref{luminosity}). It is about a factor
of two less than the X-ray luminosity of SS~Cyg at optical maximum
\citep[which corresponds to the second, weaker peak in the X-ray light
curve of SS Cyg;][]{whea}, and it is only a factor of two more
luminous than RU~Peg in outburst, and a factor of three brighter than
SU~UMa in outburst \citep{bas05}. GW~Lib seems to stand out due to its
unusually low X-ray luminosity in quiescence, rather than an
exceptional X-ray luminosity in outburst. It may be that the
relatively high mass of the white dwarf \citep{tow04} accounts for the
high outburst X-ray luminosity, as it may do also in SS~Cyg.

Very few systems have good X-ray coverage during an outburst, so it
is not clear whether the steep decline in the X-ray luminosity of
GW~Lib during the first ten days of outburst is typical of other
systems. The X-ray flux of WZ~Sge itself does decline during the first
half of the outburst \citep{whe05}, but only about a factor of three
in ten days, compared with a factor of thirty in GW~Lib also over ten
days (although note that the first X-ray observation of WZ~Sge
occurred about 3 days after the optical maximum). SS~Cyg also declined
after the optical maximum, in this case by about a factor of six over
seven days \citep{whea}. In contrast, the X-ray flux of VW~Hyi was
approximately constant during the outburst \citep{whe96b}.

In most dwarf novae the dominant high-energy emission during an
outburst is optically-thick emission from the boundary layer, which
emerges in the extreme-ultraviolet \citep{pri77}. This seems also to
be the case for GW~Lib, with a supersoft component detected in the
first {\it Swift} observation, although the luminosity of this
component is poorly constrained by our observations (Fig.\ref{bolom}).
The supersoft component is detected only in our first observation, but
it is likely to be present also in the later epochs and just too soft
to be detected by the {\it Swift} XRT. Only a small spectral change
would be needed to move this component out of our bandpass. Indeed,
the extreme-ultraviolet components of VW~Hyi and WZ~Sge are not
detected at all with the {\it ROSAT} PSPC and the {\it Chandra} ACIS
X-ray detectors respectively \citep{whe96b,whe05}, although they are
detected in the {\it EXOSAT} LE and {\it Chandra} LETG bandpasses.

In the few cases where good coverage has been achieved, the
extreme-ultraviolet emission rises only after the X-ray emission has
been suppressed \citep[e.g.][]{whe96a,jon92,whea}, and it is assumed
that this supersoft component takes over as the main source of cooling
as the boundary layer becomes optically-thick to its own emission.
Since the extreme-ultraviolet emission is present even in our first
{\it Swift} observation, it is possible that the observed peak in
X-ray emission actually represents the suppressed level, and that an
even stronger X-ray peak was missed, corresponding to the peak
emission of the optically-thin boundary layer. In SS~Cyg this first
X-ray peak is a factor of three brighter than the second, weaker peak
corresponding to the optical maximum.

In our later {\it Swift} observations the X-rays decline, then flatten
off, and at the end of the disc outburst there is a sharp dip followed
by a bump in the X-ray light curve, which coincides with the rapid
decline in the optical and ultraviolet lightcurves. The X-ray
hardness also increases at this time. These features are shared to
some extent with other systems. A dip and a bump are seen in U~Gem
\citep{mat00} which is the only other system where X-rays are known to
be brighter in outburst than in quiescence. A bump is also seen in
SS~Cyg where it is thought to correspond to the boundary layer
transitioning back to its optically-thin state \citep{whea}. Another
feature similar to GW Lib is the increase in hardness at the end of
the outburst in SS~Cyg, and indeed, dwarf novae are usually harder in
quiescence than in outburst \citep{bas05}.

When comparing the outburst X-ray emission to quiescence it is
important to distinguish between pre- and post-outburst quiescence.
GW~Lib was unusually faint for a dwarf nova in quiescence in the {\it
XMM-Newton} observation made two years before the 2007 outburst
\citep{hil}. Our {\it Swift} XRT outburst observations continued for
about six days after the end of the sharp decline in the optical and
ultraviolet lightcurves, which presumably defines the end of the disc
outburst. Our measured X-ray luminosity after this decline (S5 in
Table~\ref{luminosity}) is a factor of fifty higher than the
pre-outburst quiescent level \citep{hil}. Our follow-up {\it Swift}
observations in 2008 and 2009 show that GW~Lib declined by a factor of
five after the outburst, but that it remained an order of magnitude
brighter than the pre-outburst observations for at least twenty one
months after the outburst. Another important difference between the
{\it XMM-Newton} and {\it Swift} observations is that \citet{hil}
found an oxygen abundance enhanced by at least a factor of six above
the solar value, whereas we find that our first outburst spectrum is
inconsistent with such a high value and that the iron abundance
appears to be significantly sub-solar. It is difficult to understand
how observed abundances can change so much between quiescence and
outburst. 

It has been noted in other systems that the X-ray flux tends to
decrease between outbursts. Examples include VW Hyi \citep{vdw87} and
SS Cyg \citep{mcg04}. This is in contrast to the usual predictions of
the disc instability model \citep[e.g.][]{las01} in which the
accretion rate gradually increases during quiescence as the disc
refills. The inferred decrease in the quiescent X-ray flux in GW Lib
is by a much larger factor than in VW Hyi and SS Cyg, but the
inter-outburst interval is also much larger in GW Lib (decades
compared with weeks and months), so there is more time for this
decrease to progress.

\subsection{Possible disc models}
The long inter-outburst intervals of GW~Lib and other WZ~Sge type
stars mean that the opportunities to study the outbursts of these
objects in detail have been very scarce. In this respect, our data
represent a rare insight into these intriguing dwarf novae. 

To date, the physical cause of the long inter-outburst times has
remained elusive. It is not at all clear why the accretion discs in
these stars should behave any differently from those in other DNe with
very similar system parameters. Yet, while the majority of
non-magnetic, short period DNe exhibit outbursts every few weeks or
months, the WZ~Sge stars outburst every few years or decades. There
are two main sets of models which attempt to explain this stark
difference in recurrence time. In order to suppress the onset of
regular outbursts and hence lengthen the inter-outburst interval,
either the quiescent viscosity must be much lower than in other
systems \citep{sma93,how95} or the inner disc must be somehow
truncated \citep{war96,mat07}. While the low viscosity models are
appealing in that they neatly explain the long recurrence times, they
remain unsatisfying in requiring the viscosity in some quiescent discs
to be different from others while, at the same time, being very
similar during outbursts. Models which appeal to inner disc
truncation, suppress regular outbursts by removing the inner region of
the accretion disc where outbursts are most easily triggered. Often
disc truncation is explained by the propeller action of the torque
exerted on the accretion disc caused by a magnetic field anchored on a
rapidly rotating primary star \citep{war96,mat07} \citep[a white dwarf
magnetic field strength of B $\sim$ 10$^5$ G was assumed for WZ Sge
by][]{mat07}. In this case, mass would accumulate at large radii
leaving a truncated and stabilised (with respect to frequent DNe
outbursts) outer disc which acts as a large reservoir of mass. If the
same physical mechanism was responsible for the long inter-outburst
timescales of all of the WZ~Sge stars, it may be reasonable to expect
their outbursts to look very similar. In this respect, the differences
between the observed outburst properties of GW~Lib and WZ~Sge, as
outlined above, are puzzling.

The detailed emission physics of the magnetic propeller models in
particular is not well-understood, making theoretical predictions of
multiwavelength outburst lightcurves extremely difficult. However, we
note that both the low viscosity and disc truncation mechanisms tend
to reduce the accretion rate during quiescence and may explain the low
and decreasing X-ray flux in GW~Lib between outbursts. Also, in the
case of GW Lib, the X-rays are quenched on a timescale of $\sim$ 10
days. The only plausible timescale close to this value is the viscous
time of the accretion disc. Interpreting this as a viscous timescale,
t$_{visc}$ $\sim$ R$^{2}$/$\alpha_{H}$c$_{s}$H, we obtain an
associated radius of R $\sim$ 10$^{10}$ cm , where we have assumed that
the viscosity in the hot state $\alpha_{H} = 0.1$, sound speed c$_{s}$
= 10 km s$^{-1}$, and disc scale height = 0.1 R. This estimate is
interesting as it is close to the required values for disc
truncation. Thus it is conceivable that the quenching of the X-ray flux
is associated with the inward progression of the accretion disc toward
the white dwarf, and the eventual development of a boundary layer,
once the outburst has been triggered.

\section{Conclusions}
We have obtained optical, UV and X-ray observations of the 2007
outburst of the WZ Sge type dwarf nova GW Lib. GW Lib stands out as
the second known dwarf nova, in addition to U Gem, where hard X-rays
are not suppressed during outburst. Rather than having a remarkably
high X-ray luminosity in an outburst, GW Lib has a very low X-ray
luminosity in quiescence compared to other dwarf novae. The outburst
X-ray lightcurve of GW Lib shows some similarities with other dwarf
novae, such as a bump seen at the end of the X-ray lightcurve and
hardening of the X-rays towards the end of the outburst. These
features are also seen in SS Cyg. WZ Sge and GW Lib show some
differences in their outburst data: the hard X-rays in WZ Sge are
suppressed and the X-rays decline with a much smaller factor in the
beginning of the X-ray lightcurve compared to GW Lib.

A supersoft component, which probably originates from the
optically-thick boundary layer, is detected in the first outburst
spectrum. Other systems, such as VW Hyi and WZ Sge also show this
component in their outburst data. The spectral resolution of the {\it
Swift} XRT or UVOT is not sufficient to distinguish emission or
absorption lines in the spectra.

The outburst X-ray luminosity at the optical maximum was three orders
of magnitude higher than during the pre-outburst quiescence level in
2005. GW Lib was still an order of a magnitude brighter during the
2008 and 2009 post-outburst observations than during the pre-outburst
observations.

The long recurrence time and the lack of normal outbursts suggest that
the structure of the accretion disc could be explained by models which
favor very long recurrence times. The two main categories for these
models are: 1) low disc viscosity in quiescence and 2) a truncated
inner disc due to a magnetic propeller white dwarf. Assuming that the
outbursts of all WZ Sge stars would be driven by a similar physical
mechanism, the observed differences in the outburst data of GW Lib and
WZ Sge are perplexing.

\section*{Acknowledgments}
KB acknowledges funding from the European Commission under the Marie
Curie Host Fellowship for Early Stage Research Training SPARTAN,
Contract No MEST--CT--2004--007512, University of Leicester, UK. The
authors also acknowledge the support of STFC. We acknowledge with
thanks the variable star observations from the AAVSO International
Database contributed by observers worldwide and used in this
research. We thank the {\it Swift} science team and planners for their
support of these target of opportunity observations. {\it Swift} data
were extracted from the {\it Swift} science archive at
www.swift.le.ac.uk. This work made use of data supplied by the UK {\it
Swift} Science Data Centre at the University of Leicester. The WASP
Consortium consists of astronomers primarily from the Queen's
University Belfast, Keele, Leicester, The Open University, and St
Andrews, the Isaac Newton Group (La Palma), the Instituto de
Astrof{\'i}sica de Canarias (Tenerife) and the South African
Astronomical Observatory. The SuperWASP-N and WASP-S Cameras were
constructed and operated with funds made available from Consortium
Universities and the UK's Science and Technology Facilities
Council. WASP-South is hosted by the South African Astronomical
Observatory (SAAO) and we are grateful for their support and
assistance. We thank M.R. Goad for helpful comments on this paper.

\label{lastpage}
\end{document}